\documentclass[11pt]{article}
\usepackage{appb,  latexsym, epsfig}

\begin{document}
\begin{titlepage}
%
\renewcommand{\thefootnote}{\fnsymbol{footnote}}
\begin{flushright}
BN-TH-98-15\\
hep-th/9807102\\
July 1998
\end{flushright}
\vspace{1cm}
 
\begin{center}
{\Large {\bf Algebraic Renormalization of}} \\[4mm]
{\Large{\bf the Electroweak Standard Model}} 
\footnote{Presented at the DESY Zeuthen Workshop on Elementary
    Particle Theory ``Loops and Legs in Gauge Theories'', Rheinsberg,
    Germany, April 19-24, 1998.}\\
{\makebox[1cm]{  }       \\[1.5cm]
{\bf Elisabeth Kraus}
\\ [3mm]
{\small\sl Physikalisches Institut, Universit\"at Bonn} \\
{\small\sl Nu\ss allee 12, D--53115 Bonn, Germany} 
} 
\vspace{1.5cm}
\vfill
{\bf Abstract}
\end{center}
\begin{quote}
The algebraic method of renormalization is applied to the standard model
of electroweak interactions. We present the most important modifications
compared to theories with simple groups. 
\end{quote}
\vfill
\renewcommand{\thefootnote}{\arabic{footnote}}
\setcounter{footnote}{0}
\setcounter{page}{0}
\end{titlepage}
\clearpage

\title{ALGEBRAIC RENORMALIZATION OF \\ THE ELECTROWEAK STANDARD MODEL
}
\author{Elisabeth Kraus
\address{Physikalisches Institut, Universt\"at Bonn, Nu\ss allee 12,
D-53115 Bonn}
}
\maketitle
\begin{abstract}
The algebraic method of renormalization is applied to the standard model
of electroweak interactions. We present the most important modifications
compared to theories with simple groups. 
\end{abstract}
  
\newcommand{\Cal}{{\cal C}}
\renewcommand{\l}{\lambda}
\renewcommand{\a}{\alpha}
\renewcommand{\b}{\beta}
\renewcommand{\d}{\delta}
\renewcommand{\k}{\kappa}
\newcommand{\ld}{\buildrel \leftarrow \over \d}
\newcommand{\rd}{\buildrel \rightarrow \over \d}
\newcommand{\e}{\eta}
\renewcommand{\o}{\omega}
\newcommand{\dem}{\d_\o}
\newcommand{\p}{\partial}
\newcommand{\pmu}{\p_\mu}
\newcommand{\pmo}{\p^\mu}
\newcommand{\pnu}{\p_\nu}
\newcommand{\s}{\sigma}
\renewcommand{\r}{\rho}
\newcommand{\bpsi}{\bar\psi}
\newcommand{\dslash}{\p\llap{/}}
\newcommand{\pslash}{p\llap{/}}
\newcommand{\ve}{\varepsilon}
\newcommand{\uvi}{\underline{\varphi}}
\newcommand{\vi}{\varphi}
\newcommand{\ue}{u^{(1)}}
\newcommand{\Am}{A_\mu}
\newcommand{\An}{A_\nu}
\newcommand{\Fmnu}{F_{\mu\nu}}
\newcommand{\Fmno}{F^{\mu\nu}}
\newcommand{\Ga}{\Gamma}
\newcommand{\Gao}{\Gamma^{(o)}}
\newcommand{\Gae}{\Gamma^{(1)}}
\newcommand{\Gacl}{\Gamma_{cl}}
\newcommand{\Gagf}{\Gamma_{{\rm g.f.}}}
\newcommand{\Gainv}{\Gamma_{\rm inv}}
\newcommand{\pvi}{\partial\varphi}
\newcommand{\om}{{\bf w}}
\newcommand{\mn}{\mu\nu}
\newcommand{\Tmn}{T_{\mn}}
\newcommand{\Gmn}{\Ga_{{\mn}}}
\newcommand{\hT}{\hat T}
\newcommand{\emt}{energy-mo\-men\-tum ten\-sor}
\newcommand{\eit}{Ener\-gie-Im\-puls-Ten\-sor}
\newcommand{\Tc}{T^{(c)}}
\newcommand{\Tcr}{\Tc_{\rho\sigma}(y)}
\newcommand{\ha}{{1\over 2}}
\newcommand{\dalam}{{\hbox{\frame{6pt}{6pt}{0pt}}\,}}
\newcommand{\wtm}{{\bf W}^T_\mu}
\newcommand{\nnp}{\nu'\nu'}
\newcommand{\mnp}{\mu'\nu'}
\newcommand{\emn}{\eta_{\mn}}
\newcommand{\mpm}{\mu'\mu'}
\newcommand{\tbw}{\tilde{\bf w}}
\newcommand{\tw}{\tilde w}
\newcommand{\hw}{\hat{\bf w}}
\newcommand{\bw}{{\bf w}}
\newcommand{\bW}{{\bf W}}
\newcommand{\ubW}{\underline{\bW}}
\newcommand{\hmn}{h^{\mn}}
\newcommand{\gmn}{g^{\mn}}
\newcommand{\ga}{\gamma}
\newcommand{\Gf}{\Gamma_{\hbox{\hskip-2pt{\it eff}\hskip2pt}}}
\newcommand{\T}{\buildrel o \over T}
\newcommand{\Lf}{{\cal L}_{\hbox{\it eff}\hskip2pt}}
\newcommand{\np}{\not\!\p}
\newcommand{\lp}{\partial\llap{/}}
\newcommand{\ah}{{\hat a}}
\newcommand{\han}{{\hat a}^{(n)}}
\newcommand{\hak}{{\hat a}^{(k)}}
\newcommand{\dv}{{\d\over\d\vi}}
\newcommand{\zze}{\sqrt{{z_2\over z_1}}}
\newcommand{\zez}{\sqrt{{z_1\over z_2}}}
\newcommand{\Hmn}{H_{\(\mn\)}}
\newcommand{\hfrac}[2]{\hbox{${#1\over #2}$}}  
\newcommand{\smdm }{\underline m \p _{\underline m}}
 \newcommand{\tsmdm }{\underline m \tilde \p _{\underline m}} 
 \newcommand{ \Wh }{{\hat {\bf W}}^K}
\newcommand{ \CS }{Callan-Symanzik}
\newcommand{ \G}{\Gamma}
\newcommand{ \bl }{\b _ \l}
\newcommand{ \kdk }{\k \p _\k}
\newcommand{\mdm }{m \p _m} 
\newcommand{ \te }{\tau_{\scriptscriptstyle 1}}
\newcommand{ \mhi }{m_H}
\newcommand{ \mf }{m_f}
\newcommand{\bare}{^o}
\newcommand{\Pol}{{\mathbf P}}
\newcommand{\brs}{{\mathrm s}}
\newcommand{\cw}{\cos \theta_W}
\newcommand{\cws}{\cos^2 \theta_W}
\newcommand{\sw}{\sin \theta_W}
\newcommand{\sws}{\sin^2 \theta_W}
\newcommand{\cg}{\cos \theta_G}
\newcommand{\sg}{\sin \theta_G}
\newcommand{\cwg}{\cos (\theta_W- \theta_G)}
\newcommand{\swg}{\sin (\theta_W- \theta_G)}
\newcommand{\cv}{\cos \theta_V}
\newcommand{\sv}{\sin \theta_V}
\newcommand{\cvg}{\cos (\theta_V- \theta_G)}
\newcommand{\svg}{\sin (\theta_V- \theta_G)}
\newcommand{\fsc}{{e^2 \over 16 \pi ^2}}
\newcommand{\cvt}{\cos \Theta^V_3}
\newcommand{\svt}{\sin \Theta^V_3}
\newcommand{\cvf}{\cos \Theta^V_4}
\newcommand{\svf}{\sin \Theta^V_4}
\newcommand{\cgt}{\cos \Theta^g_3}
\newcommand{\sgt}{\sin \Theta^g_3}



\newcommand{\Identity}{\scalebox{1}[.95]{1}\hspace{-3.5pt}{\scalebox{1}[1.1]{1}}}
\newcommand{\slashed}{\hspace{-1.1ex}/}
\newcommand{\Slashed}{\hspace{-1.7ex}/}
\newcommand{\equ}[1]{\begin{gather} #1 \end{gather}}
\newcommand{\equa}[1]{\begin{align} #1 \end{align}}
\newcommand{\derivative}[2]{\ensuremath{\frac{\text{d} #1}{\text{d} #2}} }
\newcommand{\pderivative}[2]{\ensuremath{\frac{\partial #1}{\partial #2}} }
\newcommand{\fderivative}[2]{\ensuremath{\frac{\delta #1}{\delta #2}}}
\newcommand{\m}{\mu}
\newcommand{\n}{\nu}
\newcommand{\g}{\gamma}
\newcommand{\tI}{\tilde{I}}
\newcommand{\RE}{\mbox{Re }}
\newcommand{\IM}{\mbox{Im }}
\newcommand{\Gm}{\Gamma^{mass}_{Dirac}}
\newcommand{\pa}{{\partial}}
\newcommand{\rra}{\ \longrightarrow \ }
\newcommand{\lla}{\ \longleftarrow\ }
\newcommand{\intd}{\int \! \mbox{d}^4 x}
\section{Introduction}
The Standard Model  (SM) of electroweak interactions has been tested to high
accuracy with the precision experiments at the Z-resonance at LEP
 \cite{datasum96}.
With  these experiments the SM in its perturbative
formulation has been tested also beyond the tree approximation. For
this reason an extensive calculation of 
 1-loop processes and also 2-loop processes has been carried out in the
past years (see \cite{hollik} for a review and references therein)
 and compared to the experimental results. A careful
analysis shows that the theoretical 
predictions  and the experiments are in excellent agreement with each
other
\cite{YeRep}. 
A necessary prerequisite for carrying out precision tests of the SM
is the consistent mathematical and physical formulation 
 in the framework of its perturbative construction. Due to the fact
that parity is broken by weak interactions, higher orders in quantum
field
theory cannot be treated by referring to an invariant regularization scheme. 
 In
ref.~\cite{KR98}
we have carried out the renormalization of the electroweak SM
to all orders by applying the algebraic method. It allows to prove
renormalizability in a scheme-independent way just by using general
properties of renormalized perturbation theory (see \cite{PISO95}
for a review to the algebraic method). The algebraic method
has been first applied to gauge theories with simple or semisimple
groups \cite{BRS75}, later on it has been extended to gauge theories with
non-semisimple
groups with several $U(1)$-factors \cite{BABE78}. The results obtained
therein are only
partially applicable to the SM  due to the fact, that 
all particles are assumed to be massive,
 but it gives a complete discussion
of  anomalies in gauge theories with non-semisimple groups.

In the  section 2 of the paper we  outline the procedure of
algebraic renormalization  and present
the basic ingredients of the method,
scheme dependence of counterterms and the quantum action principle.
  In section 3 we discuss the  characterization of invariant counterterms
to the SM Green functions. There we direct our attention to three important
results: The construction of the abelian Ward identity, the definition
of  symmetry operators in the
on-shell schemes and consequences of rigid symmetry in the gauge-fixing and
ghost sector of the action.


\section{The algebraic method of renormalization}

Starting point for the construction is the BRS-symmetric classical
action of the electroweak SM. (We do not include
 strong interactions in the present analysis, but assume the quarks
 to be color vectors of global $SU(3)$.) It consists of the
$SU(2) \times U(1)$-gauge invariant action $\Ga_{GSW}$ and the
BRS-invariant gauge-fixing $\Ga _{g.f.}$ and ghost action $\Ga_{ghost}$.
\begin{equation}
\label{Gacl}
\Gacl = \Ga_{GSW} +  \Ga_{g.f.} + \Ga_{ghost} \quad
\mbox{with} \quad \brs \Gacl = 0\;.
\end{equation}
Here $\brs $ denotes the nilpotent BRS-transformations ($\brs ^2  = 0$).
The  Glashow-Salam-Weinberg action $\Ga_{GSW}$
includes the massive gauge bosons of weak interactions,
$W_\pm, Z$, and the massless photon
$A_\mu$, the leptons $ e, \nu^L_e$, the quarks $u,d$, the physical Higgs
$H$ and the unphysical scalar bosons $\phi_\pm$ and $\chi$. 
Masses of gauge bosons, fermions and the Higgs are generated by
spontaneous breaking of gauge symmetry to the electromagnetic subgroup.
We do not consider mixing of different fermion families
and assume
 CP-invariance throughout.
Free parameters of the model are the masses 
and one coupling constant, which in the QED-like
parametrization is chosen to be the electromagnetic
coupling.
For ensuring renormalizability and off-shell infrared existence
by power counting 
we choose the  restricted linear $R_\xi$ gauge in the
tree approximation. 
It  contains two gauge parameters $\xi $ and $\zeta$ and
 turns out  to be compatible with
rigid symmetry:
\begin{eqnarray}
\label{F_a}
& \Ga_{g.f.} = \intd \bigl( \frac \xi 2 B_a \tilde I_{ab} B_b +
B_a \tilde I_{ab} F_b \bigr)\;, &  \\
& F_{\pm}\equiv \partial_{\mu}W^{\mu}_{\pm} \mp
iM_{W}\zeta\phi_{\pm}, \quad
F_Z \equiv  \partial_{\mu}Z^{\mu}-M_{Z}\zeta{{\chi}}, \quad
F_A \equiv   \partial_{\mu}A^{\mu}.   & \nonumber
\end{eqnarray}
The gauge-fixing action breaks  gauge invariance.
 Introducing
the Faddeev-Popov ghosts $c_a$ and the corresponding antighosts $\bar
c_a, a= +,-, Z,A,$  the gauge-fixing action is complemented by the
 ghost action in such a way that the complete action is BRS-invariant.

In perturbation theory the Green
functions are formally defined from the classical action by the Gell-Mann-Low
formula and by Wick's theorem or equivalently
by  the Feynman diagrams and  Feynman rules. (Feynman rules of the
SM are given in
several
publications, see e.g.~\cite{Dehab}.)
 The loop
corrections to Green functions are plagued with divergencies, which have
to be consistently removed in the procedure of renormalization. 
Then  one has to prove that the symmetries of the tree
approximation can be established in the course of renormalization
 and that these symmetries uniquely fix the
 Green functions, if one imposes a finite number of normalization
conditions. 

For renormalization only the 1PI Green functions are relevant, which
are summarized in the functional of 1PI Green functions
$\Ga$. The lowest
order of $\Ga$ is the classical action,
$\Ga = \Gacl + O(\hbar) $.
For proceeding to higher orders
the  symmetry transformations have to be rewritten  into  functional form.
  The functional form of BRS-symmetry is the
Slavnov-Taylor (ST) identity.
By introducing an external scalar doublet $\hat \Phi$ we are
able to maintain rigid $SU(2)$-symmetry and 
 spontaneously broken $U(1)$-gauge symmetry for the special 
choice of gauge parameters (\ref{F_a}) and to establish the respective
Ward identities:
\begin{eqnarray}
\label{funid}
& {\cal S} (\Gacl) = 0\;, 
\qquad \qquad {\cal W}_\a \Gacl = 0 \/ , \, \a = +,-,3
 & \nonumber \\
 & \Bigl(\frac e \cw {\mathbf w}_4^Q - \sin \theta_W
{\delta \over \delta  Z^\mu} - \cos \theta_W {\delta \over \delta  A^\mu} 
\Bigr) \Gacl = \Box (\sw B_Z + \cw B_A )\;.  
& 
\end{eqnarray}
The local Ward identity is crucial for the unique
construction of higher orders.  
 The symmetry operators depend explicitly on
the free parameters of the theory. 
With the gauge choice (\ref{F_a})
they depend
on the mass ratio $M_W \over M_Z$ in the tree approximation \cite{KR98}.
We now assume that we have already calculated the 1-loop order in
a specific scheme of renormalization and denote the finite Green functions
by $\Ga^R$.
Regardless of special properties of the scheme we can
apply the action principle in its quantized version to the Green
functions \cite{LAM73},
in order to get information of the possible breakings at 1-loop order.
Applied to the symmetries of the tree approximation it tells that
in the 1-loop order the symmetries are at most broken by {\it local}
field polynomials with a definite UV and IR degree  of power counting.
Taking  the most important example, the ST identity, it reads
\begin{eqnarray}
\label{QAP}
 {\cal S} (\Ga^ R )  ^{(\leq 1)}
 = \Delta_{brs}^{(1)} \/;
& \quad &
\dim^{UV}\! \Delta^{(1)} _{brs} \leq 4 \/,\quad  \dim^{IR}\! \Delta^{(1)} 
_{brs} \geq 3 \/.   
\end{eqnarray}
The breakings include  in a first step all local field polynomials
compatible with the UV and IR dimension.  They are 
restricted if one takes into account that the renormalization schemes do not
break global symmetries  such as charge conservation, and discrete symmetries
such as  CP-invariance.
 For example $\Delta^{(1)} _{brs}$ has Faddeev-Popov charge 1, is neutral with
respect to electromagnetic charge
and is  even with respect to CP-transformations.

In the next step we have to prove that the breakings of the
ST identity can be absorbed into counterterms to the classical action:
For doing this one has to note that Green functions, when they are computed
in a specific scheme, are only 
defined up to local counterterms. These counterterms are restricted
 by the global symmetries and the discrete
symmetries of the SM.  In order to maintain the properties of power counting
renormalizability their UV  and IR degree has to agree with
 the one of the classical action
\begin{equation}
\label{CT}
\Ga^{ (\leq 1)} = {\Ga^R} ^{(\le 1)}+ \Ga_{ct}^{(1)} \quad
\hbox{with} \quad
\dim^{UV} \Ga_{ct}^{(1)}  \le 4 \/, \quad \dim^{IR} \Ga_{ct}^{(1)}  \ge 4  \;.
\end{equation}

Combining both the quantum action principle (\ref{QAP})
and the scheme dependence of
counterterms (\ref{CT}) we get
\begin{equation}
\label{ST1loop}
{\cal S}(\Ga) = {\cal S}(\Ga^R + \Ga^{(1)}_{ct}) + O(\hbar ^2) =
\Delta_{brs}^{(1)} + \brs_{\Gacl} \Ga_{ct}^{(1)} + O(\hbar ^2)\;,
\end{equation}
and a similar expression for rigid Ward identities.
Eventually we have to prove that breakings of the ST identity 
can be written as $\brs_{\Gacl}$-variations of counterterms
to the classical action, i.e.
\begin{equation}
\label{BRSCT}
\Delta_{brs}^{(1)} \stackrel ! = - \brs _{\Gacl} \Ga_{ct}^{(1)}\;.
\end{equation}

Up to this point we did only use properties of renormalized perturbation
theory. Finally one has to characterize  both the counterterms
and the breakings in terms of the symmetries:
 First, counterterms  have to be decomposed  into
invariant and non-invariant counterterms
\begin{equation}
\Ga_{ct} = \Ga_{inv} + \Ga_{break} \qquad \hbox{with} \qquad
\brs_{\Gacl}
\Ga_{inv} = 0\;.
\end{equation}
The coefficients of  invariants are not determined by the symmetries
but  have to be fixed by normalization conditions. 
Second, one restricts the breakings by using algebraic
properties
of the symmetry operators, as e.g.~nilpotency
of the ST operator:
\begin{eqnarray}
\label{brsnil}
&\brs _\Gamma \,{\cal S}
 (\Gamma) = 0  \qquad \hbox{and} \qquad 
\brs _\Gamma \, \brs_\Gamma  = 0 \quad \hbox{if} \quad {\cal S}(\Gamma)
=0  \;.
\end{eqnarray}
Applying  the $\brs_{\Gacl}$-operator to eq.~(\ref{ST1loop}) 
one obtains from (\ref{brsnil})
that $\Delta^{(1)}_{brs}$ is  $\brs_{\Gacl}$-invariant
\begin{equation}
\brs_{\Gacl}
 \Delta_{brs}^{(1)} = 0 \;.
\end{equation}
$\Delta ^{(1)} _{brs} $ is invariant under the $\brs_{\Gacl}$-transformation
whenever it is  a variation of the counterterms, i.e.\ 
if eq.~(\ref{BRSCT}) is fulfilled.
 If there is only one field polynomial
which is  $\brs_{\Gacl}$-invariant,
but not a variation of counterterms,
one has  an anomaly, and symmetries cannot be established by adjusting
counter\-terms. In this way one has achieved
an  algebraic characterization of scheme dependent breakings
and anomalies. The proof to all orders proceeds by induction, passing through
the same steps as above from order $n$ to $n+1.$

In the SM the algebraic characterization of breakings can be proven to be the
same as 
in the symmetric $SU(2) \times U(1)$ theory and
does  not depend on the specific form of spontaneous symmetry
breaking. In \cite{BABE78} 
it has been shown that    there are only the well-known Adler-Bardeen
anomalies.
Their coefficients vanish in the SM, if we include lepton and colored quark
doublets.
 The difficult and indeed specific
part 
 is 
 the classification of invariant
counterterms and of  appropriate normalization conditions.
Here mixing effects between neutral massless/massive fields have to
be carefully analysed from the point of view of off-shell infrared existence.
In the next section we present three important results of this analysis.


\section{Invariant counterterms and normalization conditions}

 Invariant counterterms are  determined if one
 solves the ST identity and the Ward identities for the most
general action compatible with power counting renormalizability:
\begin{equation}
\label{gagencl}
{\cal S}(\Ga_{cl}^{gen}) = 0 \qquad {\cal W}_\a\Ga_{cl}^{gen} = 0
\qquad \dim^{UV}\Ga_{cl}^{gen} \leq 4\/.
\end{equation}
In the SM
 one finds as solution an action, which contains
in addition to the  free parameters 
of the tree approximation
two further undetermined couplings in each fermion family.  They
are couplings of 
abelian currents to the abelian component of vector fields and are not
determined by the ST identity, but have to be fixed by a local
gauge Ward identity.
(In \cite{GR98} these couplings are fixed
 by an antighost 
equation. From  there   the local Ward identity  is defined by using the
consistency with the ST identity.)
 In the SM classically we have three types of abelian currents,
the
currents of lepton and quark family number  conservation, 
$ j^\mu_{l}$ and $ j^\mu_{ q} $, and the
sum of the electromagnetic and neutral 
current of weak interactions. 
Being more specific we find as a special solution of eq.~(\ref{gagencl})
\begin{equation}
\label{galq}
\Ga'_{cl}  = \Ga_{cl}
+  \intd (g_l j^\mu_{l}+ g_q
j^\mu_{q})(\sin\theta_W Z_\mu + \cos \theta_W A_\mu )  
\end{equation}
where 
$\Gacl$ is the classical action (\ref{Gacl}).
In order to identify the action (\ref{Gacl}) as solution of symmetry
identities, 
 one has to impose the local
$U(1)$-Ward identity as given in (\ref{funid}). 
 The abelian local Ward operator is to all orders fixed to be the sum of the
non-integrated neutral $SU(2)$ Ward operator and of the electromagnetic
charge operator 
\begin{equation}
\label{w4Q}
{\mathbf w}_4^Q \equiv   {\mathbf w}_{em} -{\mathbf w}_3 \qquad
\mbox{with} \quad \Bigl[{\mathbf w}_4^Q , {\cal W}_\a \Bigr] = 0
\end{equation}
 In fact
the local abelian Ward identity (\ref{funid}) 
with the abelian operator (\ref{w4Q}) is the functional
generalization  of the classical
Gell-Mann-Nishijima relation.
 We want to point out that
the abelian Ward identity has to be characterized to be abelian by its
commutation relations with Ward operators of
rigid $SU(2)$-symmetry. For this reason it is  crucial to establish
 rigid symmetry in addition to the ST identity.

We did already mention that the symmetry operators depend explicitly
on the mass ratio $M_W /M_Z $ in the tree approximation. 
Solving eq.~(\ref{gagencl}) by inserting the tree operators
  we are not able to  fix all mass parameters by  normalization
conditions and especially we are not able to diagonalize the mass
matrix of neutral vector bosons on-shell at the same  time:
\begin{eqnarray}
\label{onshell}
&\RE \Bigl. \G_{W^+W^-} \Bigr|_{p^2 = M_W^2} = 0, 
\qquad
\RE \Bigl. \G_{ZZ} \Bigr|_{p^2 = M_Z^2} = 0, 
\qquad 
\Bigl. \G_{AA} \Bigr|_{p^2 = 0} = 0, &
\nonumber\\
 & \RE \Bigl. \G_{ZA} \Bigr|_{p^2 = M_Z^2} = 0, 
\qquad 
 \Bigl. \G_{ZA} \Bigr|_{p^2 = 0} = 0. &
\end{eqnarray}
Mass diagonalization at $p^2 = 0$ is crucial for obtaining infrared finite
expression for off-shell Green functions and are implemented  in
the BPHZL scheme by the IR power counting.
In order to fulfill
 the normalization conditions (\ref{onshell}) one
has to introduce a non-diagonal wave-function renormalization for the
neutral vector bosons.
If we carry out such field redefinitions also in the symmetry operators,
the symmetry operators are renormalized and get higher order corrections, i.e.
\begin{equation}
{\cal S} (\Ga) \to  ({\cal S} + \delta {\cal S} )(\Ga)\;, \quad
{\cal W}_\a \Ga \to  ({\cal W}_\a + \delta {\cal W}_\a )\Ga \/.
\end{equation}
These corrections are in agreement with the algebra.
For this reason
 one cannot fix the symmetry operators to their tree form
but one has to take the most general ones compatible with  the algebraic
properties of the tree approximation (see (\ref{brsnil})),
 when one solves eqs.~(\ref{gagencl}).
Then   one has  indeed
the same number of free parameters as normalization
conditions even
in the complete on-shell scheme.

Requiring rigid symmetry has important restrictions on
 the gauge fixing.
As we have already mentioned the choice (\ref{F_a}) is covariant under 
rigid $SU(2) \times U(1)$-transformations and can be constructed as
being invariant under rigid transformation by introducing the external 
scalar field $\hat \Phi$. In this special gauge the ghost mass ratio
is equivalent to the vector mass ratio, 
\begin{equation} 
 {M_W^{\mathrm ghost}}  / {M_Z^{\mathrm ghost}} ={M_W}/ {M_Z} +
{\cal O} (\hbar) \/ ,
\end{equation}
but it
 turns out to be differently renormalized from the vector mass ratio in higher
orders.
For this reason we have to introduce the ghost mass ratio as an independent
parameter of the model.
However, if one takes 
arbitrary gauge parameters $\zeta_W$ and $\zeta_Z$ in the
in the gauge-fixing functions $F_\pm$
and $F_Z$  without changing $F_A $ (\ref{F_a}), 
one breaks rigid symmetry by the gauge fixing  
and 
cannot derive Ward identities of rigid and local  $U(1)$ symmetry.
Consequently we  loose the control about the gauged abelian currents in higher
orders, 
namely we are not able to identify the electromagnetic current by means
of a local Ward identity. Taking for the gauge fixing 
\begin{eqnarray}
\label{F_agen}
 & F_{\pm}\equiv \partial_{\mu}W^{\mu}_{\pm} \mp
iM_{W}\zeta_W\phi_{\pm},
 \qquad 
F_Z\equiv \partial_{\mu}Z^{\mu}-M_{Z}\zeta_Z{{\chi}},\nonumber{ } \\
&F_A\equiv   \partial_{\mu}A^{\mu} - \zeta_Z M_Z \frac \cw \sw \bigl(1-
\frac {\zeta_W} {\zeta _Z} \bigr) \chi\;,  &
\end{eqnarray}
the gauge fixing is  compatible with rigid symmetry and  allows at
the same time to treat the ghost mass ratio as an independent parameter
of the theory. In order to avoid non-diagonal ghost mass terms 
 BRS-transformations have to be generalized to
$\brs \bar c_a = \hat g_{ab} B_b  $
and $\hat g$ differs from the unit matrix by
\begin{equation}
\begin{array} {lcl}
\hat g_{ZZ}& = & \cos (\theta_G-\theta_W ) \\
\hat g_{AZ}&  = & \sin (\theta_G -\theta_W)
\end{array}
\qquad
\begin{array} {lcl}
\cos \theta_W &  = & {M_W \over M_Z}  \\
{\cos \theta_G  \over  \cos (\theta_W - \theta_G)}& = & {\zeta_W M_W \over
\zeta_Z M_Z} \end{array}
\end{equation}
With this modification 
 the ghost mass matrix is diagonal, and can be
achieved  to be diagonal on-shell in higher orders 
by adjusting the free parameters in the matrix $\hat g_{ab}$.
Then also the Faddeev-Popov part is free from off-shell infrared divergencies.
Contrary to simple gauge groups we have  introduced not only
independent wave function renormalizations for ghosts and vectors, but have
also an independent wave function  renormalization  for antighosts.



\end{document}